# ORIGIN OF ISOTOPIC DIVERSITY AMONG CARBONACEOUS CHONDRITES

Jan L. Hellmann[1,2,*], Jonas M. Schneider[1,3], Elias Wölfer[1,3], Joanna Drążkowska[3], Christian A. Jansen[1], Timo Hopp[1,3], Christoph Burkhardt[1,3], Thorsten Kleine[1,3]

[1]Institut für Planetologie, University of Münster, Wilhelm-Klemm-Str. 10, 48149 Münster, Germany
[2]Department of Geology, University of Maryland, 8000 Regents Drive, College Park, MD 20742, USA
[3]Max Planck Institute for Solar System Research, Justus-von-Liebig-Weg 3, 37077 Göttingen, Germany

*Correspondence to: hellmann@umd.edu

## ABSTRACT

Carbonaceous chondrites are some of the most primitive meteorites and derive from planetesimals that formed a few million years after the beginning of the solar system. Here, using new and previously published Cr, Ti, and Te isotopic data, we show that carbonaceous chondrites exhibit correlated isotopic variations that can be accounted for by mixing among three major constituents having distinct isotopic compositions, namely refractory inclusions, chondrules, and CI chondrite-like matrix. The abundances of refractory inclusions and chondrules are coupled and systematically decrease with increasing amount of matrix. We propose that these correlated abundance variations reflect trapping of chondrule precursors, including refractory inclusions, in a pressure maximum in the disk, which is likely related to the water ice line and the ultimate formation location of Jupiter. The variable abundance of refractory inclusions/chondrules relative to matrix is the result of their distinct aerodynamical properties resulting in differential delivery rates and their preferential incorporation into chondrite parent bodies during the streaming instability, consistent with the early formation of matrix-poor and the later accretion of matrix-rich carbonaceous chondrites. Our results suggest that chondrules formed locally from isotopically heterogeneous dust aggregates which themselves derive from a wide area of the disk, implying that dust enrichment in a pressure trap was an important step to facilitate the accretion of carbonaceous chondrite parent bodies or, more generally, planetesimals in the outer solar system.

## 1. INTRODUCTION

Meteorites are fragments of planetesimals formed within the first few million years (Myr) of the Solar System and as such provide a unique window into the early evolution of the solar protoplanetary disk. Based on their isotopic composition, meteorites can be subdivided into the non-carbonaceous (NC) and carbonaceous (CC) groups (Budde et al. 2016a; Warren 2011), which represent two spatially distinct disk reservoirs that coexisted for several Myr and likely correspond to the inner and outer Solar System, respectively (Kruijer et al. 2017; Warren 2011). Preservation of the NC-CC isotopic dichotomy requires a dynamical barrier against mixing and homogenization of the two reservoirs, such as the early formation of Jupiter (Kruijer et al. 2017; Morbidelli et al. 2016), a pressure maximum in the disk (Brasser & Mojzsis 2020), likely related to the water ice line (Charnoz et al. 2021; Izidoro et al. 2022; Lichtenberg et al. 2021; Morbidelli et al. 2022), or a combination thereof.

Carbonaceous chondrites are eponymous for the CC group of meteorites and primarily consist of refractory inclusions, chondrules, and metal embedded into a fine-grained matrix (*e.g.,* Scott & Krot 2014). The abundances of these components vary among the different carbonaceous chondrite groups, where CI (Ivuna-type) chondrites are comprised almost entirely of matrix, while other chondrites such as the CV (Vigarano-type) chondrites contain abundant chondrules and refractory inclusions including Ca-Al-rich inclusions (CAIs) and amoeboid olivine aggregates (AOAs). These components are chemically and isotopically distinct, and their heterogeneous distribution can account for the chemical and isotopic variations observed among bulk carbonaceous chondrites (*e.g.,* Alexander 2019; Hellmann et al. 2020). However, the dynamical processes that led to the variable distribution of these components and how these processes are related to the accretion of carbonaceous chondrite parent bodies are unclear. For instance, it has been suggested that refractory inclusions formed close to the Sun, were subsequently transported into the outer disk (*e.g.,* Ciesla 2007; Cuzzi et al. 2003), and then possibly trapped for ~1-3 Myr in a pressure maximum just outside Jupiter's orbit where the carbonaceous chondrite parent bodies later formed (Desch et al. 2018; Scott et al. 2018). However, whether similar processes also affected the other constituents of chondrites (*i.e.,* chondrules and matrix), and whether these components were mixed before or during parent body accretion is unclear (Gerber et al. 2017; Schneider et al. 2020; Van Kooten et al. 2021).

Here we use variations in the Ti, Cr, and Te isotopic compositions of carbonaceous chondrites to assess the dynamical processes that led to the variable distribution of chondrite components among carbonaceous chondrites. This in turn is key for understanding the dynamics of material transport through the disk accompanying the formation of some of the most pristine planetesimals of the Solar System. Whereas the Ti and Cr isotope anomalies are of nucleosynthetic origin and ultimately reflect the heterogeneous distribution of presolar components, the Te isotope variations result from mass-dependent fractionation during physicochemical processes in the accretion disk. Unlike previous studies, we focus on ungrouped or anomalous carbonaceous chondrites, which are compositionally distinct from the major carbonaceous chondrite groups and, hence, may have formed from distinct precursor materials.

## 2. SAMPLES AND METHODS

The samples of this study include ten ungrouped or anomalous carbonaceous chondrites, which are compositionally and/or texturally distinct from the known major chondrite groups as well as the CR1 chondrite GRO 95577, which has a unique O isotopic composition and an anomalously high matrix abundance (Schrader et al. 2011). This sample set is complemented by two additional ungrouped carbonaceous chondrites for which combined Te-Cr-Ti (Flensburg; Bischoff et al. 2021) and Te isotopic data (Tagish Lake; Hellmann et al. 2020) were reported previously. For Tagish Lake, new Cr and Ti isotopic data were obtained in this study. In addition, new Cr and Ti isotopic data are also reported for the CI chondrite Orgueil; this sample was included because of its important role in defining the composition of primitive matrix in carbonaceous chondrites (*e.g.,* Alexander 2019). Importantly, taking the ungrouped and anomalous chondrites into account more than doubles the number of distinct parent bodies from the CC reservoir for which combined Te, Cr, and Ti isotope data are available and allows for a comprehensive assessment of the nature and extent of isotopic variations within this group of meteorites.

All samples (>1 g except for ~0.4 g for Orgueil) were powdered, and separate aliquots were taken for Te (70-100 mg) and Ti–Cr (15-25 mg) isotope analyses. Samples for Te measurements were mixed with a $^{123}$Te-$^{125}$Te double spike, dissolved in HF–$HNO_3$ on a hotplate, and Te was purified by ion-exchange chromatography (Hellmann et al. 2021; Hellmann et al. 2020). Samples for Ti and Cr isotope analyses were digested in Parr bombs with 6 mL 1:1 HF–$HNO_3$ at 190 °C for 4 days and then repeatedly dissolved and dried in HCl–$HNO_3$. Titanium and Cr were separated from the sample matrix by ion-exchange chromatography as outlined in Gerber et al. (2017) and Schneider et al. (2020). The Te and Ti isotope measurements were performed using a ThermoScientific Neptune *Plus* multi-collector inductively coupled plasma mass spectrometer following established measurement routines (Gerber et al. 2017; Hellmann et al. 2020). The Cr isotope measurements were conducted on a ThermoScientific Triton *Plus* thermal ionization mass spectrometer following the protocol of Schneider et al. (2020), but using a multi-static instead of a single line static data acquisition routine. The mass-dependent Te isotope variations are reported as $\delta^{128/126}$Te values (per mil deviation from NIST SRM 3156), and nucleosynthetic Ti and Cr isotope anomalies are reported as μ-values (parts-per-$10^6$ deviations from the NIST SRM 3112a and the Origins Lab OL-Ti standards) after mass-bias correction by internal normalization to $^{49}$Ti/$^{47}$Ti=0.749766 (Niederer et al. 1981) and $^{50}$Cr/$^{52}$Cr=0.051859 (Shields et al. 1966) (Table 1).

## 3. CORRELATED ISOTOPIC VARIATIONS AMONG CARBONACEOUS CHONDRITES

### 3.1. Mixing of Chondrules/Chondrule Precursors with CI Chondrite-like Matrix

The ungrouped and anomalous carbonaceous chondrites of this study together with samples from the major groups of carbonaceous chondrites plot on a single $\delta^{128/126}$Te–1/[Te] correlation line (Figure 1). For the major groups of carbonaceous chondrites, $\delta^{128/126}$Te is also correlated with the mass fraction of matrix, indicating that the correlated variations of $\delta^{128/126}$Te and Te content reflect variable amounts of CI chondrite-like matrix (Hellmann et al. 2020). This in turn implies that chondrules are characterized by light Te isotopic compositions (*i.e.,* low $\delta^{128/126}$Te), which is approximately the same for the major carbonaceous chondrite groups. The low $\delta^{128/126}$Te of the chondrules may result from Te isotope fractionation during chondrule formation but may also be inherited from the precursor material of the chondrules (see Hellmann et al. 2020 for details). For the samples of this study, the volume fractions of matrix are known from the petrographic classification of the meteorites, but since the bulk densities of the samples are unknown, these volume fractions cannot easily be translated into matrix mass fractions. Thus, it is difficult to assess whether these samples also plot on a correlation line between $\delta^{128/126}$Te and matrix mass fraction, but we note that the elevated $\delta^{128/126}$Te values of these samples are consistent with their generally high matrix volume fractions.

Hellmann et al. (2020) showed that for CV, CO, CM, and CI chondrites, as well as for Tagish Lake, $\delta^{128/126}$Te is also correlated with nucleosynthetic $\mu^{54}$Cr anomalies (Figure 1b). These correlations likely reflect mixing between volatile-rich, isotopically heavy (*i.e.,* higher $\delta^{128/126}$Te), and $^{54}$Cr-rich CI chondrite-like matrix with volatile-poor and isotopically light chondrules/chondrule precursors. This is consistent with the observation that chondrules from CV, CO, and CM chondrites, despite the large variations among individual chondrules from each group, are characterized by the same average $\mu^{54}$Cr of ~60 (Hellmann et al. 2020). This indicates that for a given chondrite group the offset of a bulk chondrite's $\mu^{54}$Cr from the mean $\mu^{54}$Cr of its chondrules also correlates with the amount of matrix in each chondrite. The new Te and Cr isotope data reveal that most ungrouped and anomalous carbonaceous chondrites also plot on the $\mu^{54}$Cr–$\delta^{128/126}$Te mixing line defined by the major groups (Figure 1b), indicating that the Cr and Te isotope variability among most carbonaceous chondrites can be accounted for by variable abundances of the same two components, namely chondrules/chondrule precursors and CI chondrite-like matrix.

The only carbonaceous chondrites clearly deviating from this general trend are the CR chondrites, which plot well below the $\mu^{54}$Cr–$\delta^{128/126}$Te mixing line defined by other carbonaceous chondrites. The CR1 chondrite GRO 95577 also plots below this mixing line but has a distinctly higher $\delta^{128/126}$Te value and Te content compared to the characteristic composition of CR chondrites (Figure 1). Of note, GRO 95577 plots on the $\delta^{128/126}$Te–1/[Te] mixing line between chondrules and matrix, indicating that its higher Te content is due to a higher portion of CI-like matrix compared to other CR chondrites and not solely the result of Te redistribution during parent body alteration. In the $\mu^{54}$Cr–$\delta^{128/126}$Te diagram, both CR chondrites and GRO 95577 plot on a mixing line between CR chondrules and CI chondrite-like matrix (Figure 1b), suggesting that the distinct isotopic composition of CR chondrites and GRO 95577 compared to other carbonaceous chondrites predominantly reflects the incorporation of a distinct population of chondrules with more elevated $\mu^{54}$Cr values compared to chondrules in CV, CM, and CO chondrites (Figure 1b). This interpretation is consistent with a model in which CR chondrules formed by recycling of an earlier chondrule generation after mixing with CI chondrite-like dust (Marrocchi et al. 2022).

The three ungrouped C2 chondrites Essebi, Tarda, and Acfer 094 also appear to plot slightly below the $\mu^{54}$Cr–$\delta^{128/126}$Te mixing line between CV-CM-CO chondrules and CI chondrite-like matrix, although this offset is not clearly resolved. If it exists, the small offset may reflect a slightly more elevated average $\mu^{54}$Cr of the chondrules in these chondrites compared to chondrules in CV, CO, and CM chondrites. As for the CR chondrites, this higher $\mu^{54}$Cr may reflect a slightly higher fraction of CI chondrite-like material among the precursors of these chondrules.

In summary, the Cr and Te isotopic variability among carbonaceous chondrites can be accounted for by mixing between chondrules/chondrule precursors and CI chondrite-like matrix. Whereas the latter appears to be contained in all carbonaceous chondrites, the $^{54}$Cr isotopic composition of the chondrule component may vary among the individual groups. This is most obvious for CR chondrites, whose chondrules have substantially more elevated $\mu^{54}$Cr values compared to chondrules from most other carbonaceous chondrites but may also account for small deviations from the $\mu^{54}$Cr–$\delta^{128/126}$Te mixing line that may exist for some ungrouped carbonaceous chondrites.

## 3.2. Correlated Abundances of Refractory Inclusions and Chondrules

The $\mu^{54}$Cr anomalies among the carbonaceous chondrites are not only correlated with $\delta^{128/126}$Te but are also inversely correlated with $\mu^{50}$Ti (Figure 2). This observation holds for the major groups of carbonaceous chondrites as well as for the ungrouped and anomalous samples of this and prior studies. This $\mu^{50}$Ti–$\mu^{54}$Cr correlation may reflect mixing between CI-like matrix and an isotopically distinct dust component characterized by elevated $\mu^{50}$Ti and low $\mu^{54}$Cr, but such a component has yet not been identified in carbonaceous chondrites (Alexander 2019). Instead, components with elevated $\mu^{50}$Ti typically also have elevated $\mu^{54}$Cr, such as for instance CAIs and AOAs, which are characterized by both positive $\mu^{50}$Ti ~900 (Davis et al. 2018; Render et al. 2019; Torrano et al. 2019; Trinquier et al. 2009) and $\mu^{54}$Cr ~600 (Birck & Allegre 1988; Larsen et al. 2011; Papanastassiou 1986). Moreover, given that CAIs are strongly enriched in refractory elements like Ti, they exert a strong control on the Ti isotopic composition of bulk carbonaceous chondrites (Trinquier et al. 2009), but contain too little Cr to significantly affect a carbonaceous chondrite's $\mu^{54}$Cr. To assess the effect of CAI (and AOA) admixture on the Ti and Cr isotopic compositions of bulk carbonaceous chondrites more quantitatively, we have calculated mixing lines between various chondrule-matrix mixtures and refractory inclusions (Figure 2b). Note that Ti isotope data are only available for chondrules from the CV chondrite Allende (Gerber et al. 2017) and chondrules in other carbonaceous chondrite groups may have different $\mu^{50}$Ti signatures. Nevertheless, the calculations reproduce the observed abundances of refractory inclusions in the various chondrite groups reasonably well, such as for instance ~2-4% CAIs/AOAs in CM chondrites and ~3-8% in CV and CO chondrites, respectively (*e.g.,* Ebel et al. 2016; Fendrich & Ebel 2021; Kimura et al. 2020). Together, these calculations show that the variations in $\mu^{50}$Ti among bulk carbonaceous chondrites are almost entirely governed by the abundance of refractory inclusions, while variations in $\mu^{54}$Cr predominantly reflect the relative proportions of chondrules and matrix (Figure 2b). A corollary of these observations is that the co-variation of $\mu^{50}$Ti and $\mu^{54}$Cr among carbonaceous chondrites appears to reflect coupled variations in the abundances of refractory inclusions and chondrules, which systematically decrease from a high abundance in CV and CO chondrites to essentially zero in CI chondrites.

## 4. ACCRETION OF CARBONACEOUS CHONDRITE PARENT BODIES

### 4.1. Transport and Mixing of Dust Before Chondrite Parent Body Accretion

Whereas CAIs are the oldest dated solids of the Solar System and, together with AOAs, are thought to have formed close to the Sun (*e.g.,* Wood 2004), most chondrules in carbonaceous chondrites formed ~2–4 Myr later (*e.g.,* Budde et al. 2016b; Kurahashi et al. 2008) and further away from the Sun. The linked abundances of refractory inclusions and chondrules, therefore, require a dynamical process that led to a coupled enrichment of two distinct dust components originating from different regions of the accretion disk into the narrow formation location of specific carbonaceous chondrite parent bodies. Furthermore, the ~2–4 Myr age difference between CAIs and chondrules implies that CAIs were stored in the accretion disk for at least this period of time before they were incorporated into their host chondrites (*e.g.,* Ciesla 2007; Cuzzi et al. 2003). Because objects the size of CAIs are expected to be rapidly lost into the Sun by gas drag, it has been suggested that the inward drift of CAIs was blocked by a "pressure trap", *i.e.,* a pressure maximum in the disk, which is also the location in the disk where carbonaceous chondrite parent bodies later formed (Brasser & Mojzsis 2020; Desch et al. 2018; Scott et al. 2018). This pressure maximum may have formed through the formation of Jupiter's core and the associated opening of a gap in the disk (*e.g.,* Kruijer et al. 2017), but other origins of this structure are also possible (*e.g.,* Izidoro et al. 2022). Regardless of its exact origin, a pressure trap provides an efficient mechanism not only for blocking the inward drift of CAIs (and AOAs), but also for producing local enrichments of these objects. This is consistent with the dearth of refractory inclusions in non-carbonaceous chondrites and their enhanced abundance in some carbonaceous chondrites (Desch et al. 2018). However, storage of refractory inclusions alone cannot account for the systematic co-variation of CAI/AOA and chondrule abundances relative to matrix, which we propose reflects the trapping of not only refractory inclusions but also chondrules or their precursors in the pressure maximum. This would naturally result in coupled variations of CAIs/AOAs and chondrules, because both components became enriched by the same process (Figure 3).

The coupled trapping of refractory inclusions and chondrules raises the question of why the abundances of these components vary among the different groups of carbonaceous chondrites. Provided that all carbonaceous chondrites formed in the same pressure maximum, this requires that the enrichment of refractory inclusions and chondrules in this structure varied in time and/or space. This could be caused by the varying supply of these components due to differences in their radial drift speed caused by their distinct aerodynamical properties defined by their grain size and internal density. Models of planetesimal formation via the streaming instability show that this mechanism prefers grains of high Stokes numbers, that is, the largest and/or densest dust aggregates (Bai & Stone 2010). This may lead to preferential incorporation of refractory inclusions and chondrules over matrix dust into chondrite parent bodies as long as these more refractory components are available. This is consistent with the observation that the parent bodies of the matrix-rich CI and CM chondrites tend to have younger inferred accretion ages than those of matrix-poor chondrites such the CV and CO chondrites (Figure 4). The only carbonaceous chondrite group plotting off this trend are the CR chondrites, which are characterized by a relatively young accretion age of ~3.6 Myr after CAI formation (*e.g.,* Budde et al. 2018; Schrader et al. 2017) but have the lowest matrix mass fraction among the carbonaceous chondrites (Figure 4). There is some uncertainty in the matrix mass fraction of CR chondrites, because different members of this group have somewhat variable matrix fractions and metal abundances. The latter exerts a strong control on the conversion of matrix volume to mass fractions and accounts for at least some of the variability in reported matrix fractions. Nevertheless, even when using the higher CR matrix mass fraction reported by Patzer et al. (2022), the CR chondrites plot off the trend of matrix mass fraction versus accretion age (Figure 4). This offset may indicate that the CR chondrites formed from distinct precursor material than the other carbonaceous chondrites or that most of their CI chondrite-like material has been processed into a second generation of chondrules (Marrocchi et al., 2022).

Despite the good correlation between accretion age and matrix content, it is important to recognize that particularly for the matrix-rich chondrites some inconsistencies exist in the inferred accretion ages. For instance, carbonates in the matrix-rich carbonaceous chondrite Flensburg and in some samples from asteroid Ryugu (which resemble CI chondrites) have $^{53}$Mn-$^{53}$Cr ages of ~ <1.8-2.6 Ma after CAI formation (Bischoff et al. 2021; McCain et al. 2023). This implies parent body accretion before that time and, hence, earlier than inferred for the matrix-rich CI and CM chondrites. Given that both Flensburg and Ryugu are characterized by similarly high matrix fraction as these two groups of chondrites, it appears that they would not plot on the trend of accretion age versus matrix content shown in Figure 4. Clearly, more work is needed to better establish the accretion ages of especially the matrix-rich carbonaceous chondrites and to more reliably assess as to whether they formed later than their matrix-poor counterparts.

### 4.2. Origin of Chondrules and their Precursors

Based on their chemical and isotopic composition, several studies have argued that the accretion region of carbonaceous chondrites is in the outer Solar System, beyond the orbit of Jupiter (*e.g.,* Kruijer et al. 2017; Warren 2011). Some other studies, however, proposed that chondrules (*e.g.,* Williams et al. 2020) and perhaps even their host carbonaceous chondrites (*e.g.,* Van Kooten et al. 2021) formed in the inner Solar System. This proposal is based primarily on the observation that some chondrules from CV chondrites have $^{54}$Cr isotopic compositions overlapping with those of NC chondrites, which are thought to represent the inner disk. Moreover, while the cores of some CV chondrules are characterized by $^{54}$Cr deficits (*i.e.,* the characteristic composition of the NC reservoir), their rims display $^{54}$Cr excesses (*i.e.,* the characteristic composition of the CC reservoir). This has been interpreted to indicate that CV chondrules in general formed in the inner disk, were subsequently transported outwards and were mixed with CI chondrite-like dust, which drifted inward towards the Sun (Van Kooten et al. 2021).

In this model, refractory inclusions would be added to the accretion location of carbonaceous chondrites together with CI chondrite-like dust by inward transport from the outer disk, while chondrules would be added by outward transport from the inner disk. However, there is no reason why in this scenario the abundances of refractory inclusions and chondrules should be coupled, because the outward-drifting region in a vicinity of the pressure trap is much narrower than the inward-drifting outer disk and it gets depleted much faster (*e.g.,* Pinilla et al. 2012). Thus, the linked abundances of refractory inclusions and chondrules argue against the massive outward transport of carbonaceous chondrite chondrules inferred based on NC-like $^{54}Cr$ signatures in some of these chondrules. Instead, these signatures, and the overall $^{54}Cr$ variability among chondrules, are best understood as reflecting isotopic heterogeneities in their precursor dust (Schneider et al. 2020). Within this framework, NC-like $^{54}Cr$ isotopic compositions of chondrules indicate that their precursor material—but not the chondrules themselves—originated in the inner disk. Similar to CAIs, this precursor material was likely transported outwards during the early viscous expansion of the disk (Nanne et al. 2019), and later drifted back towards the Sun by gas drag. This is consistent with chemical and isotopic evidence showing that CAIs have partially been incorporated into chondrules (*e.g.,* Gerber et al. 2017), indicating that CAIs were part of the chondrule precursor material and that, therefore, the coupled abundances of CAIs and chondrules were established prior to chondrule formation.

## 5. CONCLUSIONS

The ubiquitous presence of multi-element isotope correlations among carbonaceous chondrites, including the major groups as well as ungrouped and anomalous samples, demonstrates that these correlations are a fundamental feature of the accretion region of carbonaceous chondrite parent bodies. These correlations indicate (1) that carbonaceous chondrites represent mixtures of three main components having distinct isotopic compositions, namely chondrules/chondrule precursors, refractory inclusions (CAIs and AOAs), and CI chondrite-like matrix, and (2) that the abundances of refractory inclusions and chondrules are coupled and systematically vary with the amount of matrix. We suggest that these coupled abundance variations reflect the trapping of chondrule precursors, including CAIs and AOAs, in a pressure maximum in the disk, which is likely related to the water ice line and the ultimate formation location of Jupiter. The variable ratio of refractory inclusions/chondrules relative to matrix is the result of their distinct aerodynamical properties resulting in differential delivery rates and their preferential incorporation into meteorite parent bodies during the streaming instability. Chondrules formed from isotopically heterogeneous precursor dust aggregates trapped in the pressure bump, suggesting that the enrichment of chondrule precursor dust in a pressure trap may have been an important step to facilitate the accretion of carbonaceous chondrite parent bodies or, more generally, planetesimals in the outer Solar System.

**ACKNOWLEDGEMENTS –** We gratefully acknowledge the constructive review by Conel Alexander and efficient editorial handling by Faith Vilas. We thank J. Zipfel (Senckenberg) for providing a sample of Essebi. The meteorite sample GRO 95577 was provided by NASA, which is also gratefully acknowledged. Antarctic meteorite samples are recovered by the Antarctic Search for Meteorites (ANSMET) program which has been funded by NSF and NASA, and characterized and curated by the Department of Mineral Sciences of the Smithsonian Institution and Astromaterials Curation Office at NASA Johnson Space Center. The work was funded by the European Research Council Advanced Grant 101019380 (HOLYEARTH) and by the European Union under the European Union's Horizon Europe Research & Innovation Programme 101040037 (PLANETOIDS). Views and opinions expressed are however those of the authors only and do not necessarily reflect those of the European Union or the European Research Council. Neither the European Union nor the granting authority can be held responsible for them. Funded by the Deutsche Forschungsgemeinschaft (DFG, German Research Foundation) – Project-ID 263649064 – TRR 170. This is TRR170 publication No. 182.

**Table 1.**
Mass-independent Ti and Cr and mass-dependent Te isotopic data together with Te concentrations of ungrouped and anomalous carbonaceous chondrites, GRO 95577 (CR1), and Orgueil (CI1).

| Sample | Type | Weight (mg) (Ti-Cr aliquot) | N (Ti-IC) | $\mu^{46}$Ti (±95% CI) | $\mu^{48}$Ti (±95% CI) | $\mu^{50}$Ti (±95% CI) | N (Cr-IC) | $\mu^{53}$Cr (±95% CI) | $\mu^{54}$Cr (±95% CI) | Weight (mg) (Te aliquot) | Te (ng/g) (±2σ) | N (Te-DS) | $\delta^{128/126}$Te (±2 s.d.) |
|---|---|---|---|---|---|---|---|---|---|---|---|---|---|
| NWA 12957 | C3.00-ung. | 24.1 | 14 | 45± 6 | 1± 4 | 286± 9 | 9 | 27± 4 | 105± 11 | 102.1 | 1527±61 | 8 | 0.12±0.02 |
| NWA 12416 | C3-ung. | 19.0 | 14 | 47± 6 | -4± 2 | 277± 10 | 8 | 14± 8 | 118± 22 | 106.4 | 1505±61 | 6 | 0.11±0.01 |
| Tagish Lake* | C2-ung. | 25.1 | 14 | 43± 6 | -2± 6 | 271± 10 | 8 | 20± 4 | 131± 13 | 78.9 | 1669±42 | 5 | 0.11±0.01 |
| Acfer 094 | C2-ung. | 19.5 | 14 | 48± 7 | 6± 6 | 275± 14 | 10 | 23± 6 | 139± 16 | 67.3 | 1574±48 | 8 | 0.08±0.02 |
| Tarda | C2-ung. | 22.0 | 14 | 44± 9 | -3± 5 | 260± 9 | 10 | 23± 7 | 140± 16 | 74.9 | 1563±42 | 5 | 0.08±0.01 |
| Essebi | C2-ung. | 21.7 | 14 | 50± 8 | -2± 3 | 286± 5 | 12 | 24± 5 | 144± 11 | 101.4 | 1406±23 | 5 | 0.06±0.02 |
| NWA 5958 | C2-ung. | 18.4 | 14 | 50± 11 | 6± 7 | 290± 12 | 10 | 19± 8 | 115± 17 | 84.6 | 1561±48 | 6 | 0.08±0.03 |
| NWA 11024 | CM-an. | 14.5 | 14 | 49± 9 | -2± 5 | 285± 9 | 6 | 24± 6 | 102± 16 | 81.1 | 1586±27 | 6 | 0.07±0.02 |
| WIS 91600 | CM-an. | 19.9 | 14 | 47± 9 | -3± 4 | 307± 5 | 6 | 14± 8 | 133± 22 | 80.8 | 1609±45 | 5 | 0.11±0.02 |
| NWA 11086 | CM-an. | 24.4 | 14 | 40± 8 | -4± 5 | 282± 9 | 4 | 14± 16 | 103± 16 | 80.6 | 1632±52 | 6 | 0.12±0.02 |
| GRO 95577 | CR1 | 25.5 | 14 | 20± 9 | -7± 5 | 199± 6 | 6 | 34± 4 | 159± 11 | 91.1 | 1151±31 | 5 | 0.02±0.03 |
| Orgueil* | CI1 | 25.0 | 12 | 31± 11 | -5± 3 | 194± 10 | 8 | 27± 5 | 173± 8 | 53.9 | 2250±63 | 5 | 0.16±0.01 |
| BIR1a | | | | | | | 14 | 6± 4 | 20± 13 | | | | |
| DTS-2b | | | | | | | 8 | 5± 2 | 8± 13 | | | | |
| JB-04 | | | 13 | -6± 7 | 1± 4 | -3± 10 | | | | | | | |
| JA-2 | | | 11 | -13± 8 | 1± 5 | -2± 10 | | | | | | | |

* Te isotopic composition and concentration have been measured previously by Hellmann et al. (2020)

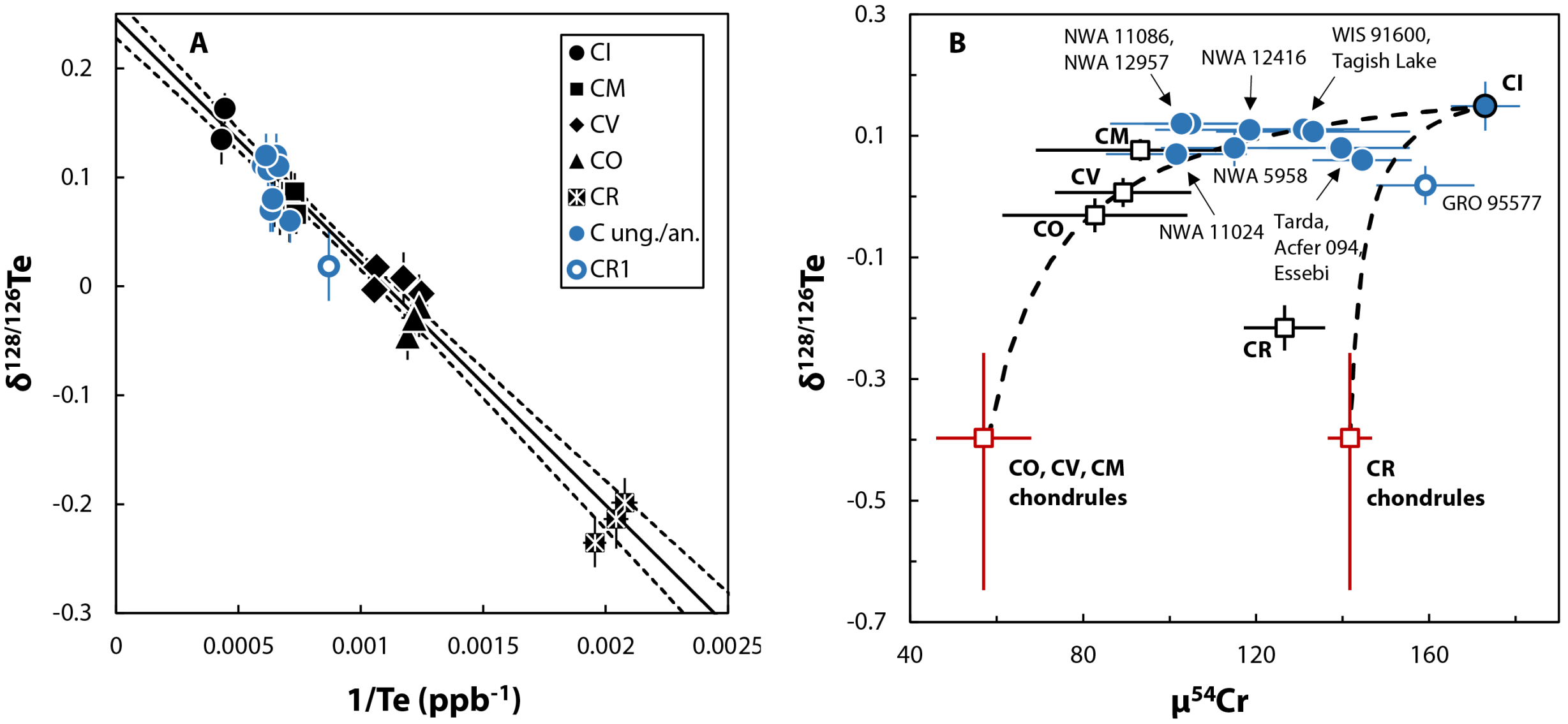

**Figure 1.** Mass-dependent Te isotope fractionation and nucleosynthetic Cr isotope anomalies of carbonaceous chondrites. **A)** Tellurium isotopic compositions and concentrations of ungrouped and anomalous carbonaceous chondrites (blue). All of the samples plot on the mixing line defined by individual samples of the major carbonaceous chondrite groups (Hellmann et al. 2020). **B)** Tellurium isotopic compositions and $\mu^{54}$Cr isotope anomalies of ungrouped and anomalous carbonaceous chondrites (blue). Black squares represent the average compositions of the major carbonaceous chondrite groups and red squares represent the average compositions of CO, CV, and CM as well as CR chondrules. Dashed lines are calculated chondrule-matrix mixing lines between chondrules and CI chondrites (CI, Te = 2.28 ppm, Cr = 2620 ppm; CO-CV-CM chondrules, Te = 0.35 ppm, Cr = 3200; CR chondrules, Te = 0.35 ppm, Cr = 4320 ppm; Alexander 2019; Gerber et al. 2017; Hellmann et al. 2020; Schneider et al. 2020; Zhu et al. 2019). The $\delta^{128/126}$Te data are from Hellmann et al. (2020) and $\mu^{54}$Cr data of carbonaceous chondrites and chondrules are compiled in Tables A1 and A2, respectively.

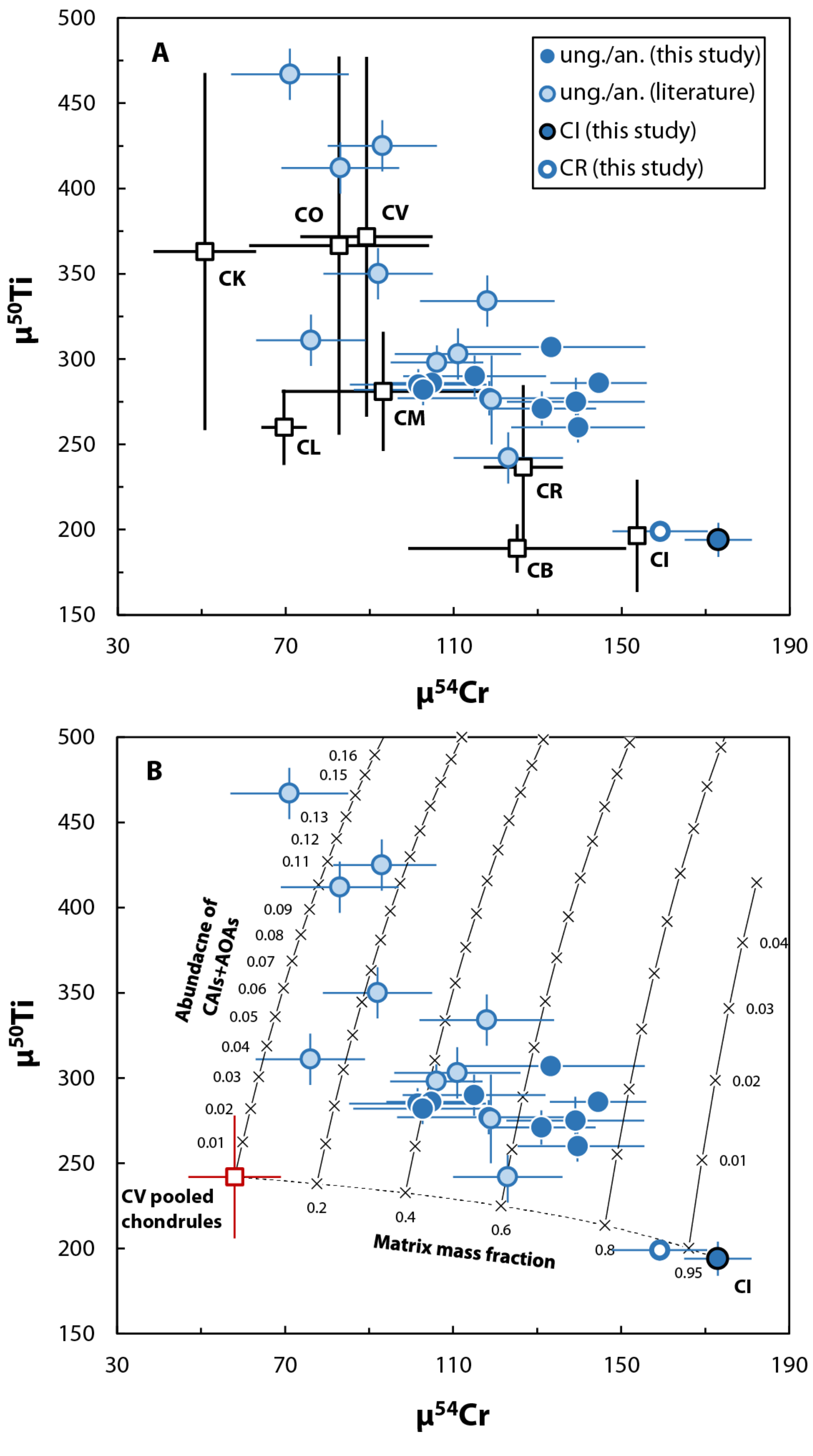

**Figure 2.** $\mu^{50}$Ti and $\mu^{54}$Cr isotope anomalies of carbonaceous chondrites. **A)** Average isotopic compositions of major carbonaceous chondrite groups (black; see Table A1), individual ungrouped and anomalous carbonaceous chondrites from this study (solid blue) and previous studies (light blue; Bischoff et al. 2021; Torrano et al. 2021), as well as of Orgueil (CI1) and GRO 95577 (CR1). **B)** Isotopic compositions of individual carbonaceous chondrites from this study and previous studies as in A). The red symbol represents the average $\mu^{50}$Ti and $\mu^{54}$Cr isotope anomalies of pooled chondrules from the CV (Gerber et al. 2017; Schneider et al. 2020)chondrite Allende (Gerber et al. 2017; Schneider et al. 2020). The mixing line between chondrules and CI chondrites (dashed line) was calculated using elemental abundances (CI chondrites, Ti = 440 ppm, Cr = 2620 ppm; chondrules, Ti = 1200, Cr = 3200) and isotopic compositions (CI chondrites, $\mu^{50}$Ti = 194, $\mu^{54}$Cr = 173; chondrules, $\mu^{50}$Ti = 242, $\mu^{54}$Cr = 58) from this and previous studies (*e.g.,* Alexander 2019; Gerber et al. 2017; Schneider et al. 2020; Zhu et al. 2019). The solid lines represent mixing lines for chondrules and refractory inclusions (CAIs and AOAs) at a given matrix mass fraction. For the refractory inclusions we used previously published isotopic data ($\mu^{50}$Ti = 900; $\mu^{54}$Cr 600; Birck & Lugmair 1988; Davis et al. 2018; Larsen et al. 2011; Papanastassiou 1986; Render et al. 2019; Torrano et al. 2019; Trinquier et al. 2009) and elemental abundances (CAIs, Ti = 5400 ppm, Cr = 200 ppm; AOAs, Ti = 2200, Cr = 2000; Komatsu et al. 2001; Trinquier et al. 2009) and a 1:1 CAI–AOA ratio (*e.g.,* Ebel et al. 2016). The CAI-to-AOA ratio in carbonaceous chondrites is not well constrained and may vary among different groups. High CAI-to-AOA ratios will result in steeper mixing lines than low CAI-to-AOA ratios.

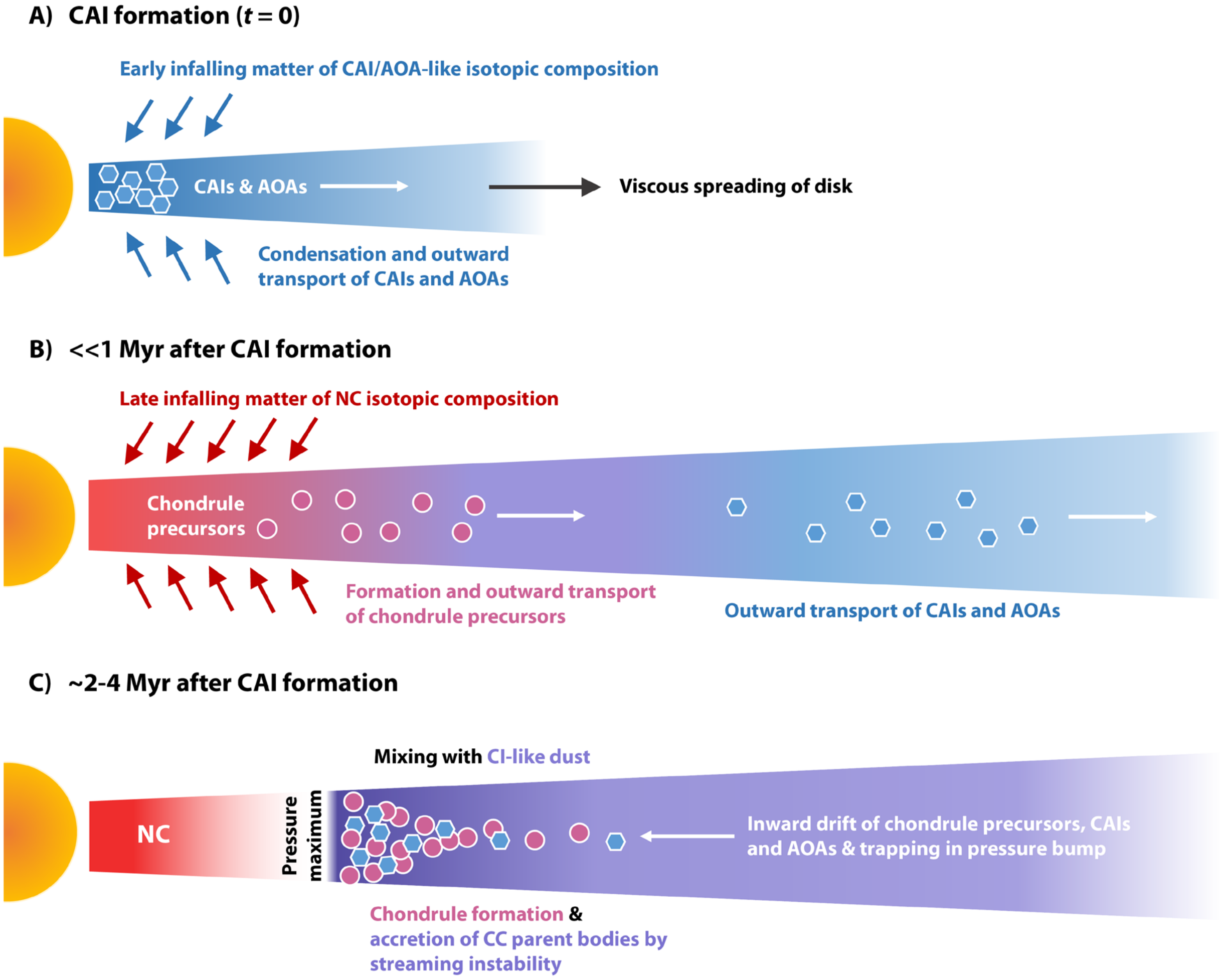

**Figure 3.** Cartoon illustrating our preferred scenario for transport and mixing of dust components before and during the accretion of the carbonaceous chondrite parent bodies. Based on model of Nanne et al. (2019) for the formation of the NC-CC dichotomy by time-varied infall from heterogenous molecular cloud core. **A)** Early infall had a CAI/AOA-like isotopic composition (blue), which reflects the isotopic composition of the early disk formed by viscous spreading. CAIs and AOAs are transported outward by the same process. **B)** Isotopic composition of the infall changes to NC-like (red), which dominates the inner disk. Chondrule precursors form in the inner disk and, like CAIs and AOAs, are transported outwards through the disk. After infall stopped, the disk exhibits an isotopic gradient from NC-like in the inner disk to CAI/AOA-like material in the outer disk. Mixing between these distinct materials in the inner disk then produced the characteristic isotopic composition of the CC reservoir. **C)** Radial drift of CAIs, AOAs, and chondrule precursors leads to their coupled enrichment in a pressure maximum, which is likely associated with the water ice line and the ultimate formation location of Jupiter. Chondrule formation and mixing of CAIs/AOAs, chondrules, and CI-like matrix occurs in this pressure trap, where the relative proportions of these components vary depending on the timing of parent body accretion.

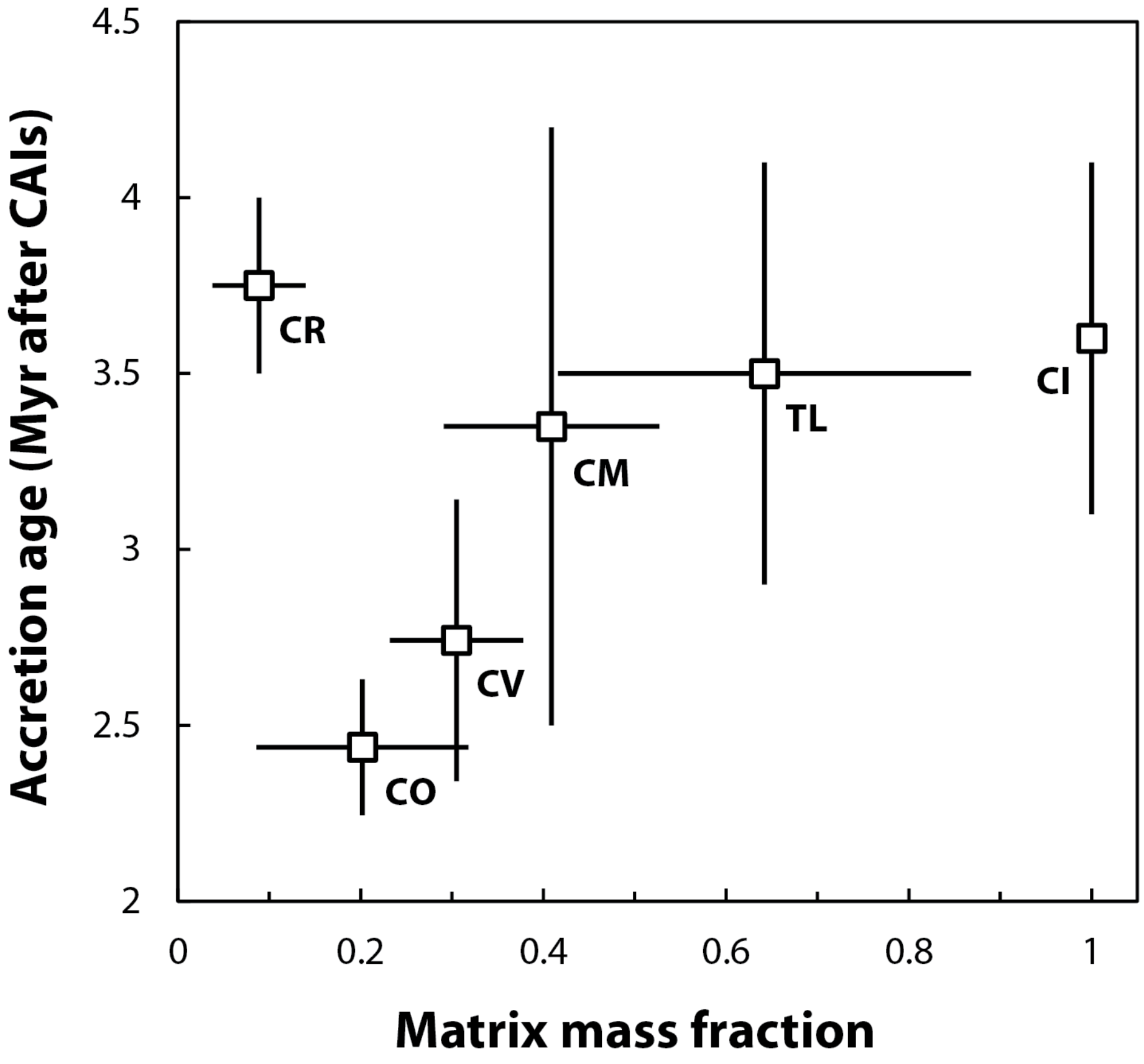

**Figure 4.** Accretion ages of carbonaceous chondrite parent bodies versus matrix mass fraction. Accretion ages are estimated based on chondrule formation ages and thermal evolution models (see Table A3 for details). Matrix mass fraction are from Hellmann et al. (2020).

## APPENDIX

**Table A1.**
Summary of Ti and Cr isotope literature data for carbonaceous chondrites

| Sample name | Type | N | $\mu^{50}Ti$ (±2σ)* | References | N | $\mu^{53}Cr$ (±2σ)* | N | $\mu^{54}Cr$ (±2σ)* | References |
|---|---|---|---|---|---|---|---|---|---|
| Orgueil | CI | 4 | 185 ± 12 | 1, 2 | 8 | 28 ± 10 | 8 | 154 ± 6 | 10-16 |
| Ivuna | CI | 1 | 208 ± 51 | 3 | 4 | 31 ± 20 | 6 | 153 ± 14 | 3, 13, 16-19 |
| Murchison | CM | 9 | 301 ± 10 | 1, 2, 4 | 6 | 23 ± 8 | 6 | 98 ± 8 | 4, 10-12, 15, 20, 21 |
| Murray | CM | 2 | 276 ± 24 | 2, 4 | 3 | 19 ± 18 | 3 | 94 ± 33 | 4, 13, 20 |
| Paris | CM | | | | 1 | 16 ± 5 | 1 | 93 ± 9 | 22 |
| Jbilet Winselwan | CM | | | | 3 | 18 ± 12 | 3 | 97 ± 27 | 16, 18, 22 |
| SCO 06043 | CM | | | | 1 | 22 ± 2 | 1 | 113 ± 12 | 16 |
| Nogoya | CM | | | | 1 | 18 ± 5 | 1 | 76 ± 4 | 16 |
| Aguas Zarcas | CM | 4 | 267 ± 57 | 4, 23 | 5 | 17 ± 7 | 5 | 95 ± 10 | 4, 16, 23 |
| Banten | CM | | | | 1 | 12 ± 3 | 1 | 86 ± 5 | 16 |
| Mighei | CM | | | | 1 | 18 ± 3 | 1 | 74 ± 10 | 20 |
| Cold Bokkeveld | CM | | | | 1 | 7 ± 3 | 1 | 81 ± 12 | 20 |
| Maribo | CM | | | | 1 | 29 ± 4 | 1 | 113 ± 15 | 20 |
| Sutter's Mill | CM | | | | 2 | 13 ± 3 | 2 | 92 ± 10 | 21 |
| NWA 8157 | CM | | | | 1 | 20 ± 11 | 1 | 101 ± 18 | 22 |
| Allende | CV | 14 | 334 ± 25 | 1, 2, 5, 6 | 16 | 11 ± 2 | 18 | 94 ± 4 | 3-5, 10-16, 24-26 |
| Bali | CV | | | | 1 | 13 ± 4 | 1 | 110 ± 6 | 16 |
| Mokoia | CV | | | | 1 | 11 ± 4 | 1 | 100 ± 1 | 16 |
| Kaba | CV | | | | 1 | 8 ± 5 | 1 | 70 ± 30 | 16 |
| Vigarano | CV | | | | 3 | 15 ± 14 | 3 | 86 ± 9 | 12, 16 |
| Leoville | CV | 1 | 409 ± 8 | 2 | 2 | 10 ± 6 | 2 | 76 ± 14 | 12, 16 |
| Felix | CO | 1 | 469 ± 12 | 1 | 1 | 7 ± 6 | 1 | 63 ± 9 | 10, 11 |
| Lance | CO | 1 | 346 ± 10 | 2 | 1 | -4 ± 7 | 1 | 57 ± 11 | 10, 11 |
| Kainsaz | CO | | | | 2 | 17 ± 10 | 2 | 95 ± 21 | 12, 13 |
| Ornans | CO | 1 | 337 ± 9 | 2 | 2 | 16 ± 10 | 2 | 97 ± 18 | 15, 16 |
| MIL 07193 | CO | | | | 1 | 16 ± 2 | 1 | 122 ± 4 | 16 |
| DOM 10104 | CO | | | | 1 | 9 ± 3 | 1 | 80 ± 6 | 16 |
| Isna | CO | 2 | 314 ± 88 | 1, 4 | 1 | 11 ± 8 | 1 | 66 ± 14 | 4 |
| Renazzo | CR | | | | 1 | 20 ± 10 | 2 | 126 ± 11 | 10, 11 |
| GRA 06100 | CR | 1 | 326 ± 9 | 7 | 2 | 26 ± 1 | 1 | 128 ± 13 | 24 |
| NWA 6043 | CR | 1 | 228 ± 10 | 8 | | | 1 | 124 ± 10 | 18 |
| EET 92161 | CR | | | | | | 1 | 119 ± 12 | 18 |
| NWA 7837 | CR | | | | | | 1 | 106 ± 8 | 18 |
| GRA 95229 | CR | 1 | 230 ± 51 | 3 | | | 1 | 118 ± 7 | 3 |
| QUE 99177 | CR | 1 | 227 ± 51 | 3 | | | 1 | 143 ± 12 | 3 |
| LAP 02342 | CR | 1 | 150 ± 51 | 3 | | | 1 | 149 ± 11 | 3 |
| Sahara 0082 | CR | 1 | 260 ± 30 | 1 | | | | | |
| NWA 801 | CR | 1 | 235 ± 4 | 2 | | | | | |
| Al Rais | CR-an | | | | 1 | 19 ± 1 | 1 | 124 ± 11 | 16 |
| Karoonda | CK | 2 | 363 ± 105 | 1, 2 | 2 | 9 ± 14 | 2 | 57 ± 18 | 10, 16 |
| ALH 85002 | CK | | | | 1 | 6 ± 2 | 1 | 46 ± 5 | 16 |
| LEW 87009 | CK | | | | 1 | 8 ± 4 | 1 | 58 ± 5 | 16 |
| EET 92005 | CK | | | | 2 | 6 ± 11 | 2 | 43 ± 27 | 12, 16 |
| Coolidge | CL | 2 | 265 ± 6 | 4, 9 | 2 | 4 ± 8 | 2 | 70 ± 31 | 4, 9 |
| Loongana 001 | CL | 1 | 257 ± 9 | 9 | 1 | 6 ± 8 | 1 | 68 ± 18 | 9 |
| LoV 051 | CL | 1 | 285 ± 8 | 9 | 1 | 5 ± 7 | 1 | 76 ± 12 | 9 |
| NWA 033 | CL | 1 | 235 ± 5 | 9 | 1 | 5 ± 11 | 1 | 64 ± 12 | 9 |
| NWA 13400 | CL | 1 | 260 ± 7 | 9 | 1 | 14 ± 13 | 1 | 70 ± 13 | 9 |
| HH237 | CB | | | | 2 | -5 ± 28 | 2 | 115 ± 78 | 13, 16 |
| Bencubbin | CB | 1 | 184 ± 43 | 1 | 2 | 4 ± 24 | 2 | 112 ± 3 | 10 |
| Gujba | CB | 1 | 194 ± 14 | 1 | 1 | -3 ± 13 | 1 | 104 ± 27 | 10 |

| | | | | | | | | |
|---|---|---|---|---|---|---|---|---|
| MIL 05082 | CB | | | 1 | 20 ± 1 | 1 | 150 ± 9 | 16 |
| QC 001 | CB | | | 1 | 19 ± 4 | 1 | 145 ± 6 | 16 |
| | | | | | | | | |
| *Average* | | | | | | | | |
| **CI** | | 2 | 196 ± 33 | 2 | 30 ± 4 | 2 | 154 ± 1 | |
| **CM** | | 3 | 281 ± 35 | 13 | 18 ± 11 | 13 | 93 ± 24 | |
| **CV** | | 2 | 372 ± 105 | 6 | 11 ± 2 | 6 | 89 ± 16 | |
| **CO** | | 4 | 367 ± 111 | 7 | 10 ± 7 | 7 | 83 ± 21 | |
| **CR** | | 7 | 237 ± 48 | 2 | 23 ± 8 | 8 | 127 ± 9 | |
| **CK** | | 1 | 363 ± 105 | 4 | 7 ± 2 | 4 | 51 ± 12 | |
| **CL** | | 5 | 260 ± 22 | 5 | 7 ± 5 | 5 | 70 ± 5 | |
| **CB** | | 2 | 189 ± 14 | 5 | 7 ± 15 | 5 | 125 ± 26 | |

*Two-sigma uncertainties represent Student’s-t 95% confidence intervals (n≥4) or two standard deviations (n<4)

**References:** (1) Trinquier et al. (2009) (2) Zhang et al. (2012) (3) Williams et al. (2020) (4) Torrano et al. (2021) (5) Burkhardt et al. (2017) (6) Sanborn et al. (2019) (7) Gerber et al. (2017) (8) Larsen et al. (2018) (9) Metzler et al. (2021) (10) Trinquier et al. (2007) (11) Trinquier et al. (2008) (12) Qin et al. (2010) (13) Shukolyukov & Lugmair (2006) (14) Petitat et al. (2011) (15) Bonnand et al. (2016) (16) Zhu et al. (2021) (17) Schiller et al. (2014) (18) Van Kooten et al. (2016) (19) Larsen et al. (2011) (20) Van Kooten et al. (2020) (21) Jenniskens et al. (2012) (22) Göpel et al. (2015) (23) Kerraouch et al. (2022) (24) Schneider et al. (2020) (25) Zhu et al. (2020a) (26) Zhu et al. (2020b)

**Table A2.**

Average Cr isotope compositions of chondrules from the major carbonaceous chondrite groups

| Sample | N | $\mu^{54}Cr$ (±95% conf.) | References |
|---|---|---|---|
| CO chondrules | 12 | 81 ± 19 | 1, 2 |
| CV chondrules | 45 | 49 ± 20 | 3, 4, 5 |
| CM chondrules | 12 | 63 ± 25 | 6 |
| **Combined CO, CV, CM chondrules** | **69** | **57 ± 14** | |
| | | | |
| CR chondrules | 31 | 142 ± 4 | 3-5, 7 |

**References:** (1) Zhu et al. (2019) (2) Qin et al. (2011) (3) Olsen et al. (2016) (4) Williams et al. (2020) (5) Schneider et al. (2020) (6) Van Kooten et al. (2020) (7) Van Kooten et al. (2016)

**Table A3.**

Accretion ages of carbonaceous chondrite parent bodies

| Type | Chondrule formation age (Myr after CAIs) | Reference | Thermal model accretion age (Myr after CAIs) | Reference | Estimated time of accretion (Myr after CAIs) |
|---|---|---|---|---|---|
| CI | | | $3.6 \pm 0.5$ | 10 | 3.1-4.1 |
| TL | | | $3.5^{+0.7}_{-0.5}$ | 10 | 3.0-4.2 |
| CM | $2.5 \pm 0.3$ | 1 | $3.5^{+0.7}_{-0.5}$ | 10 | 2.5-4.2 |
| CV | $2.2 \pm 0.5$ | 2 | $2.5 \pm 0.1$ | 11 | 2.3-3.1 |
| | $2.5 \pm 0.5$ | 3 | $3.3 \pm 0.1$ | 12 | |
| | $3.0 \pm 0.4$ | 4 | $3.0 \pm 0.2$ | 10 | |
| CO | $2.4 \pm 0.7$ | 5 | $2.3 \pm 0.15$ | 11 | 2.2-2.6 |
| | $2.5 \pm 0.3$ | 1 | $2.7 \pm 0.2$ | 10 | |
| CR | $3.6 \pm 0.6$ | 6 | $3.5 \pm 0.5$ | 10 | 3.5-4.0 |
| | $3.7^{+0.3}_{-0.2}$ | 7 | | | |
| | $3.7 \pm 0.6$ | 8 | | | |
| | $4.0 \pm 0.6$ | 9 | | | |

**References:** (1) Fukuda et al. (2022) (2) Budde et al. (2016b) (3) Nagashima et al. (2017) (4) Hutcheon et al. (2009) (5) Kurahashi et al. (2008) (6) Budde et al. (2018) (7) Schrader et al. (2017) (8) Amelin et al. (2002) (recalculated by Schrader et al. 2017) (9) Bollard et al. (2017) (10) Sugiura & Fujiya (2014) (11) Doyle et al. (2015) (12) Jogo et al. (2017)